\documentclass[preprint,showpacs,preprintnumbers,amsmath,amssymb]{revtex4}%
\usepackage{graphicx}
\usepackage{dcolumn}
\usepackage{bm}
\usepackage{amsmath}
\usepackage{amsfonts}
\usepackage{amssymb}
\begin{document}
\title{Evidence for an unfolded border-collision bifurcation in paced cardiac tissue}
\author{Carolyn M. Berger,$^{\ast\S }$ Xiaopeng Zhao,$^{\dag\S }$ David G.
Schaeffer,$^{\ddag\S }$ Hana M.
Dobrovolny,$^{\ast}$ Wanda Krassowska,$^{\dag\S }$ and Daniel J. Gauthier$^{* \dag\S }$}
\affiliation{$^{*}$Department of Physics, $^{\dag}$Department of Biomedical Engineering,
$^{\ddag}$Department of Mathematics, and $^{\S }$Center for Nonlinear and
Complex Systems, Duke University, North Carolina, 27708 USA}
\date{\today}

\begin{abstract}
We investigate, both experimentally and theoretically, the bifurcation to alternans in heart tissue. Previously, this phenomenon has been modeled either as a smooth or as border-collision period-doubling bifurcation.  Using a new experimental technique, we find a hybrid behavior: very close to the bifurcation point the dynamics are smooth-like, whereas further away they are border-collision-like. This behavior is captured by a new type of model, called an unfolded border-collision bifurcation.
\end{abstract}

\maketitle


Many nonlinear systems display a bifurcation,
where the system's response changes qualitatively as an adjustable parameter
- the bifurcation parameter - is varied \cite{strogatz}.  Bifurcation theory provides useful tools to
understand the behavior of the system in the vicinity of the bifurcation.
\ The most widely investigated nonlinear systems are those described by smooth
differential equations or maps. As a result, there exists a vast literature on the
classes of bifurcations occurring in these systems.  On the other hand,
bifurcations occurring in piecewise smooth systems, such as border-collision
bifurcations, are still under investigation and are known to display much
richer behaviors \cite{zhusubaliyev}.  Border-collision bifurcations are
believed to occur in a variety of systems, such as mechanical and electrical
devices that involve sudden switching of a component, and in economic systems
where decisions are made when a variable crosses a threshold.

The primary purpose of this Letter is to describe experimental observations and a resulting new model of
the bifurcation to alternans (defined below) in paced bullfrog cardiac muscle.
\ We find that the bifurcation mediating this transition displays both smooth
and piecewise-smooth characteristics.  Specifically, by investigating the system's
sensitivity to perturbations, we find smooth-like behavior near the bifurcation point while, further away, the behavior is border-collision like.  This apparently contradictory behavior is
placed in the context of a new type of model, which we call an unfolded
border-collision bifurcation.  The bifurcation to alternans is a crucial
problem to study because there is evidence that it can initiate ventricular
fibrillation \cite{karma1994, pastore, rosenbaum}, which often underlies
sudden cardiac death, one of the leading causes of death in the United States
\cite{aheart}.  Therefore, determining the type of bifurcation is a crucial step in 
reaching the ultimate goal of suppressing cardiac alternans \cite{chen}. 

Before describing our findings, we review briefly the behavior of paced
cardiac muscle.  An applied electrical stimulus induces an action potential,
which is the characteristic time course of the transmembrane voltage. In experiments, the
dynamical state of the tissue is often described by measuring the action
potential duration (APD).  The pacing interval, known as the basic cycle length (B),
serves as the bifurcation parameter. Under various conditions
of periodic electrical pacing, M stimuli can elicit N responses
(M:N behavior) for different values of B, where the transition from one
response pattern to another under changes in B is mediated by a bifurcation.
For slow pacing, 1:1 behavior is usually observed, where each action
potential is identical. As B shortens, a period-doubling bifurcation
sometimes occurs \cite{hall}, giving rise to a long-short periodic pattern in
APD known as a 2:2 rhythm or alternans \cite{mines}.  
\begin{figure}[ptb]
\includegraphics[width=0.75\columnwidth]{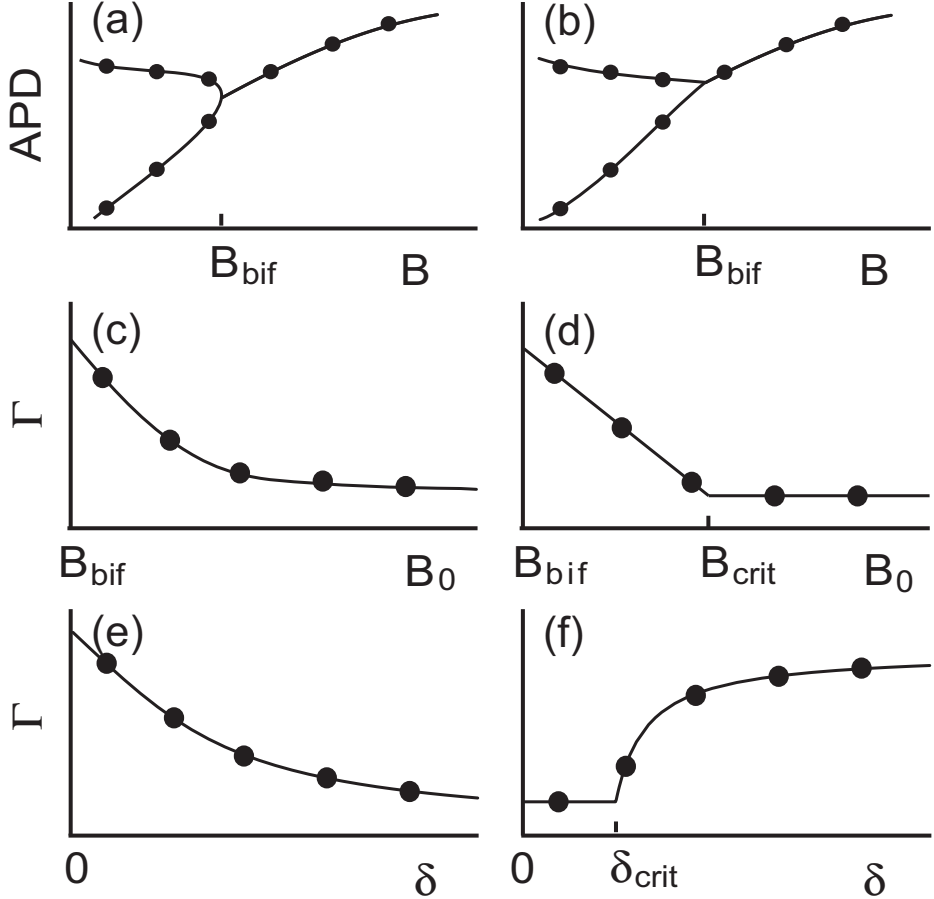}\caption{Schematic bifurcation diagrams with discrete sampling.  The sampled points (solid dots) are identical for the (a) smooth and (b) border-collision bifurcation.  (c-f) Alternate pacing: the trend in $\Gamma$ vs. B is illustrated for a (c) smooth and (d) border-collision bifurcation and the trend in $\Gamma$ vs. $\delta$ is illustrated for a (e) smooth and (f) border-collision bifurcation. }%
\label{xiaopeng_pics}%
\end{figure}

Beginning with the pioneering work of Nolasco and Dahlen \cite{nolasco}, it is
generally believed that the transition to alternans is due to a smooth
supercritical period-doubling bifurcation.  More recently, Sun \textit{et
al.} \cite{sun} introduced a model for atrioventricular nodal conduction in which it has been shown \cite{chen,hassouneh} that the transition to alternans arises from a border-collision bifurcation.  The bifurcation diagrams, where
the steady-state values of APD are plotted as a function of B, are different for these two cases.  For a
smooth period-doubling bifurcation, the bifurcated curves become tangent at
the bifurcation point, as shown in Fig. 1(a), whereas they are generally not
tangent for a border-collision bifurcation, as shown in Fig. 1(b).  Unfortunately, it is exceedingly difficult to measure the bifurcation
diagram with sufficient resolution to distinguish between these behaviors experimentally 
because only a finite number of pacing intervals can be tested before tissue damage occurs.  For example, Figs. \ref{xiaopeng_pics}(a) and (b) show that the same set of data (solid dots) is consistent with both types of models.

 We apply a new technique involving alternate pacing \cite{zhao2} to investigate the bifurcation type in cardiac tissue.  This method is a modification of a technique introduced by Heldstab \emph{et al.} \cite{heldstab} for general dynamical systems.  Specifically, this method relies on periodic perturbations to the bifurcation parameter to investigate the dynamics of the system in the 1:1 region.  More recently, Karma and Shiferaw \cite{naspe} suggest that Heldstab's method \cite{heldstab} can be exploited in a clinical setting to study the likelihood of an unfavorable cardiac event.

\ Our alternate pacing protocol is implemented in the following way.  In the 1:1 regime, we perturb the nominal pacing
interval $B_{0}$ by a small value $\delta$ so that, for the $n^{th}$ stimulus,
\begin{equation}
B_{n}=B_{0}+(-1)^n\delta.
\end{equation}
As a result, APD alternates in a long-short pattern with the corresponding
steady-state values denoted by $APD_{\text{long}}$ and $APD_{\text{short}}$.
To measure the system's sensitivity to perturbations, we define a gain as
\begin{equation}
\Gamma=\frac{APD_{\text{long}}-APD_{\text{short}}}{2\delta}. 
\label{gain}
\end{equation}

In previous papers \cite{zhao,zhao2}, we have shown that, for the two types of bifurcations, $\Gamma$ depends on the bifurcation parameter $B_0$ and perturbation amplitude $\delta$ in qualitatively different ways.
Specifically, we computed $\Gamma$ under variations of both $B_{0}$ and $\delta$. For
a smooth bifurcation, $\Gamma$ decreases monotonically as $B_{0}$ increases
compared to $B_{\text{bif}}$ [see Fig. \ref{xiaopeng_pics}(c)]. This is because the bifurcation parameter's sensitivity to
perturbations decreases further away from the bifurcation point. Likewise, $\Gamma$ decreases monotonically as $\delta$ increases [see Fig. \ref{xiaopeng_pics}(e)]. Specifically, $\Gamma$ is
proportional to $\delta^{-2/3}$ for a range of $\delta$, and it saturates for
sufficiently small $\delta$ \cite{zhao}. On the other hand, for a
border-collision bifurcation, $\Gamma$ is a piecewise smooth function of $B_{0}$ and $\delta$.  Overall, when the alternating response crosses the border (which occurs at $B_{crit}$ and $\delta_{crit}$), $\Gamma$ exhibits a decreasing trend
as $B_{0}$ increases  [see Fig. \ref{xiaopeng_pics}(d)] but an increasing trend as $\delta$ increases [see Fig. \ref{xiaopeng_pics}(f)] \cite{zhaonote}.  Although Figs. \ref{xiaopeng_pics}(c) and (d) are slightly different, it is difficult to distinguish their differences with discretely-sampled data.  However, because of significant dissimilarity between Figs. \ref{xiaopeng_pics}(e) and (f), the distinction between the two bifurcation types is evident even with just a
few data points and in the presence of experimental noise.

We apply the alternate pacing protocol in 6 adult bullfrogs. In these
experiments, the heart is excised from an adult bullfrog (Rana catesbeiana) of
either sex. After the pacemaker cells are cut away, the top half of the
ventricle is removed and placed in a chamber that is superfused with a
recirculated physiological solution \cite{hall}. A bipolar extracellular
electrode applies $2$-ms-long rectangular current pulses to the epicardial
surface of the tissue \cite{approved}. The current, whose typical amplitude
is twice the value needed to elicit a response for slow pacing, activates
a propagating excitation wave. Transmembrane voltage is measured with a glass microelectrode filled with KCl
conducting fluid. Before collecting any data, the tissue is paced at
$B_{0}=1,000$ ms for about $30$ minutes. We collect data at a sampling rate of
$4$ kHz. The data is then processed in custom-written Matlab code.

We implement the alternate pacing protocol experimentally by repeatedly carrying out the following steps:  1)  We pace at $B_{0}$ for two minutes, during which we record the
transient behavior of the APDs as they reach a steady-state of either a 1:1
or 2:2 rhythm; 2) We perturb $B_{0}$ for 20 seconds at one value of $\delta$ and record the subsequent APDs; 3) We repeat step 2 three more times, each time with a new $\delta$ value (for 11 of the 12 trials, the values of $\delta$ sweep
from high to low, to include $20$ $\mathrm{{ms}}$, $15$ $\mathrm{{ms}}$, $10$
$\mathrm{{ms}}$ and $5$ $\mathrm{{ms}}$; in one trial, we use the reverse.); and 4) The perturbations are turned off and pacing at
$B_{0}$ is resumed for 20 seconds to check that the steady-state value of the
system did not drift.  The preceding steps are repeated for decreasing values of $B_0$ until persistent alternans over
several $B_{0}$s ensue, or the cell fails to respond to every applied
stimuli.  Twelve of our trials exhibited a transition to alternans.  Typically, $B_{0}$ is decreased in steps of $25$\thinspace ms, which means that the last $B_0$ with a 1:1 response is within $25$\thinspace ms of $B_{\text{bif}}$.

\begin{figure}[ptb]
\includegraphics[width=0.95\columnwidth]{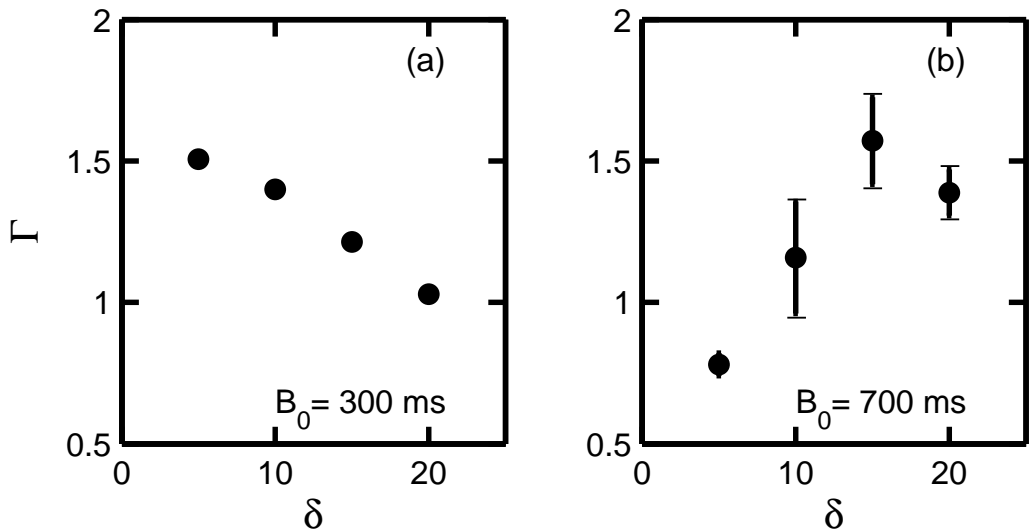}
\caption{Typical experimental results displaying two different trends in $\Gamma$ vs. $\delta$ as revealed by alternate pacing for two different frogs.  The trend is consistent with (a) a smooth period-doubling bifurcation and (b) a border-collision bifurcation. }%
\label{sm_vs_bcb_data}
\end{figure}

To analyze the results, we determine $\Gamma$
vs. $\delta$ at each $B_{0}$.  In $4$ trials from three frogs, we find that $\Gamma$ shows a
\emph{decreasing} trend as $\delta$ increases, where a typical example is shown in Fig. \ref{sm_vs_bcb_data}(a),
which agrees with a smooth bifurcation [recall Fig. \ref{xiaopeng_pics}(e)]. 
However, $4$ other trials from two frogs
demonstrate an \emph{increasing} trend in $\Gamma$ as $\delta$ increases, where a typical example is shown in Fig. \ref{sm_vs_bcb_data}(b), which agrees with a border-collision bifurcation [recall Fig. \ref{xiaopeng_pics}(f)].  In three other trials from three frogs there is no significant
variations in $\Gamma$ for different $\delta$'s and therefore cannot be
classified into either category.  The remaining trial is discussed in detail below.  For all experimental data in this paper, the error bars represent the statistical and systematic error in the measurement of APD.  In the cases where the error bars are not evident, the error is smaller than the symbol size.

\begin{figure}[ptbptb]
\includegraphics[width=0.95\columnwidth]{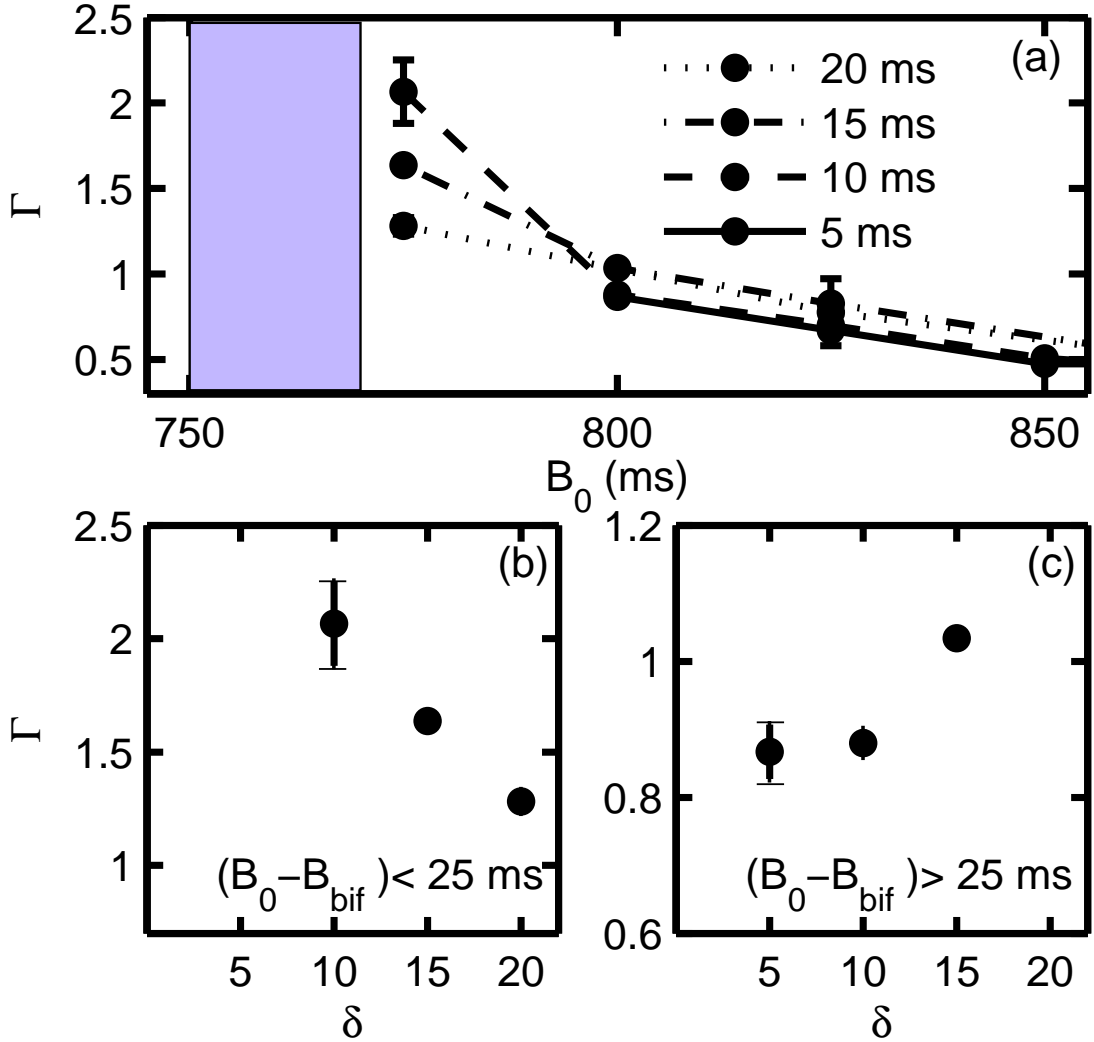}
\caption{Experimental evidence for both a smooth and border-collision bifurcation in a single trial.  (a) The trend
in $\Gamma$ vs. $B_0$ for four different values of $\delta$ (legend). Note
that the curves cross, indicating the transition from border-collision to smooth dynamics.  The bifurcation occurs somewhere in the gray area, whose precise location cannot be determined because of our finite sampling.  The same data is replotted in (b) and (c).  The behavior in (b), where $(B_0-B_{bif}) < 25~\rm{ms}$, is consistent with a smooth bifurcation and in (c), where $25~\rm{ms}< (B_0-B_{bif})< 50~\rm{ms}$, is consistent with a border-collision bifurcation.}%
\label{gain_bcls_delta}%
\end{figure}

Surprisingly, in the remaining experimental trial, we see a smooth behavior when $B_{0}%
-B_{\text{bif}}$ $\approx 25$ ms, but a border collision behavior when
$B_{0}-B_{\text{bif}}$ $\approx 50$ ms, as shown in Figs. \ref{gain_bcls_delta}(b) and (c), respectively.
 These results are unanticipated because we expected to find evidence supporting either one or the other of the two theorized bifurcation types.  Figure \ref{gain_bcls_delta}(a) shows $\Gamma$ vs. $B_{0}$ for different values of $\delta$, where it is seen that these curves cross one another.  This crossing is not indicative of either a smooth or border-collision period-doubling bifurcation, and is one of our primary experimental results.\begin{figure}[ptb]
\includegraphics[width=0.95\columnwidth]{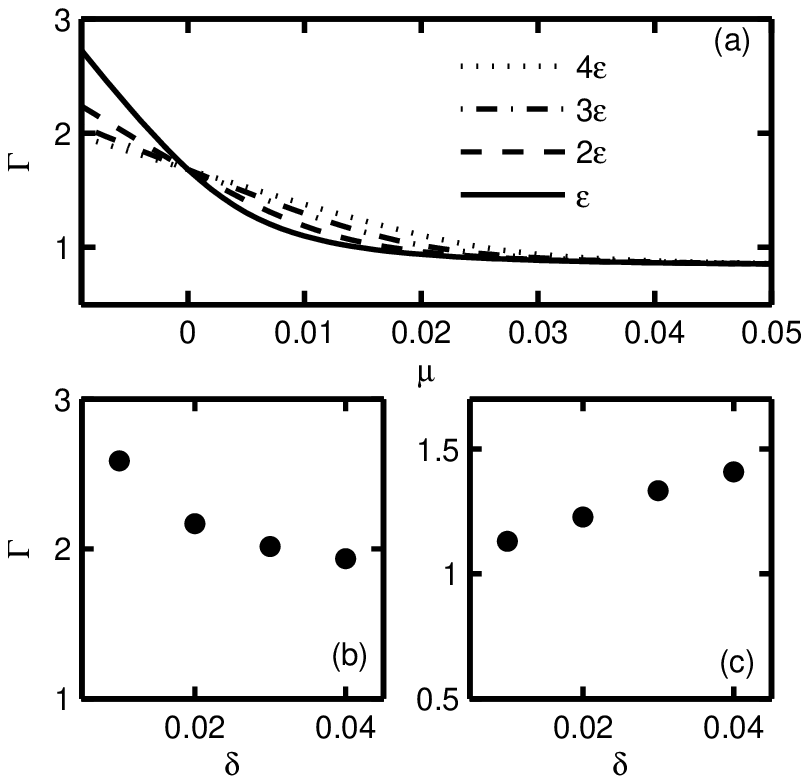}\caption{Theoretically predicted behavior of the unfolded border-collision bifurcation [map (\ref{eqn:smooth_tent})].  (a) $\Gamma$
as a function of the bifurcation parameter $\mu$ for various values of the perturbation size $\delta$ (legend, $\epsilon = 0.01$). The same data is replotted in (b) and (c).  $\Gamma$ vs. $\delta$ for (b) $\mu = -0.0073$, a distance
of 0.0018 from the bifurcation point and (c) $\mu = 0.01$, a distance of 0.0191 from the bifurcation point.  The trend in
$\Gamma$ vs. $\delta$ is consistent with a smooth bifurcation in case (b) and a border-collision bifurcation in case (c).}%
\label{new_model}%
\end{figure}

We find that these seemingly contradictory observations can be explained by a mathematical model that unfolds a border-collision
bifurcation. Consider iterations of the skewed tent map (in one dimension)
\begin{equation}
x_{n+1}=\mu-\alpha\,x_{n}-\beta\,\left\vert x_{n}\right\vert ,
\label{eqn:condense_tent}%
\end{equation}
which is singular at the border {$x=0$} in state space. This
map exhibits a border-collision period-doubling bifurcation under the
condition: $-1<\alpha+\beta<1<\alpha-\beta$ and $-1<\alpha^{2}-\beta^{2}<1$.
Replacing $\left\vert x\right\vert $ by $\sqrt{x^{2}+\varepsilon^{2}}$, where
$\varepsilon$ is a small parameter, results in an unfolding of the piecewise-smooth map (\ref{eqn:condense_tent})
\begin{equation}[floatfix]
x_{n+1}=\mu-\alpha\,x_{n}-\beta\,\sqrt{x_{n}^{2}+\varepsilon^{2}}.
\label{eqn:smooth_tent}%
\end{equation}
This new map can be viewed as an approximation of the skewed tent map
(\ref{eqn:condense_tent}). In the limit $\varepsilon\rightarrow$
$0$, map (\ref{eqn:smooth_tent}) reduces to map (\ref{eqn:condense_tent}).
Map (\ref{eqn:condense_tent}) exhibits a border-collision
period-doubling bifurcation while, for $\epsilon \neq 0$, the unfolded map (\ref{eqn:smooth_tent}) exhibits a smooth
period-doubling bifurcation. However, the dynamics of map
(\ref{eqn:condense_tent}) and map (\ref{eqn:smooth_tent}) exhibit no
significant differences except when $\mu$ is less than or on the order of $\varepsilon$. 

We investigate map (\ref{eqn:smooth_tent}) with $\alpha= 2.5$, $\beta= -2.3$, and $\epsilon= 0.01$.    The dynamical variable is $x$ (similar to APD) and the bifurcation parameter is $\mu$ (similar to $B_0$).  With this nonzero $\epsilon$, the bifurcation of map (\ref{eqn:smooth_tent}) occurs earlier than for map (\ref{eqn:condense_tent}), specifically at $\mu_{\text{bif}}=-0.0091$ rather than for $\mu_{\text{bif}}=0$ for map (\ref{eqn:condense_tent}).  We apply the alternate pacing protocol to map (\ref{eqn:smooth_tent}) by perturbing $\mu$
with $\left(-1\right)^{n}\delta$.  Figure \ref{new_model}(a) shows $\Gamma$ vs. $\mu$ for different values of $\delta$.  These curves cross one another \cite{dgs_note}.  Although the bifurcation is smooth at sufficiently small scales, it shows behavior qualitatively similar to a border-collision bifurcation when $\mu>0$, in good agreement with our experimental findings [see Fig. \ref{gain_bcls_delta}(a)].  In Figs. \ref{new_model}(b) and (c), $\Gamma$ vs. $\delta$ is plotted for $\mu= -0.00773$ and $\mu= 0.01$, respectively.  For negative values of $\mu$, $\Gamma$ decreases as $\delta$ increases, a trend consistent with a smooth bifurcation [see Fig. \ref{new_model}(b)]. On the
other hand, for positive values of $\mu$, $\Gamma$ increases as $\delta$ increases, a trend consistent
with a border-collision bifurcation [see Fig. \ref{new_model}(c)].  We note that the region over which the dynamics are smooth-like [$\mu<0$ in Fig. \ref{new_model}(a)] can be very small and on the order of $\epsilon$ in width. Therefore,
a smooth bifurcation may be masked by a border-collision bifurcation,
depending on the proximity to the bifurcation point.

Our results call to question the suggested clinical use of 
alternate pacing \cite{naspe}.  In a typical smooth bifurcation, one 
expects large $\Gamma$s far above the bifurcation point.  Hence, the 
propensity for alternans could be revealed using alternate pacing for values of $B_0$ far from the bifurcation.  It was suggested \cite{naspe} that such a procedure might be useful in the clinic because 
it would use pacing rates that are slow enough to avoid 
a potentially
 life-threatening situation.
 However, we find that $\Gamma$ remains small until the 
pacing rates are decreased to a neighborhood of the bifurcation.  Consequently, our research suggests that one would not be able to probe instabilities clinically through the use of alternate pacing.

Our experimental results reveal that the bifurcation, although smooth on a fine scale, may exhibit behavior similar to a border-collision 
bifurcation on a coarse scale. This phenomenon is described by a new model called an unfolded border-collision 
bifurcation.  We speculate that the origin of this behavior is due to rapid changes in cellular behavior during 
the course of the action potential \cite{plonsey}, such as those in the 
current-voltage or the current-concentration relation \cite{bassani}. 
\begin{center}
\textbf{Acknowledgments}
\end{center}We gratefully acknowledge the financial support of the NSF under grant PHY-0243584 and the NIH under grant 1R01-HL-72831.


\begin{thebibliography}{99}
\bibitem{zhaonote} $B_{crit}$ can be very close to $B_{bif}$, making it difficult to detect the decreasing trend in $\Gamma$ as $B_0$ increases.  Similarly, $\delta_{crit}$ can be very small, making the saturated part to the left of $\delta_{crit}$ in Fig. \ref{xiaopeng_pics}(f) hard to observe.  See \cite{zhao2} for details.
\bibitem{dgs_note} For the precise normal form described by map (\ref{eqn:smooth_tent}), these curves all cross one another at the point $\mu = 0$, but this behavior is not robust under modifications to map (\ref{eqn:smooth_tent}).
\end{thebibliography}
\end{document}